\documentclass[prd,aps,floatfix,notitlepage,nofootinbib,letterpaper,reprint,
]{revtex4-1}

\pdfoutput=1

\usepackage{graphicx}% Include figure files
\usepackage{dcolumn}% Align table columns on decimal point
\usepackage{bm}
\usepackage{amsmath}
\usepackage{amssymb}
\usepackage{hyperref}
\usepackage{color}
\usepackage{epsfig}
\usepackage{aas_macros}
\usepackage{multirow}
\usepackage{verbatim}
\usepackage{url}

\newcommand{\mr}{\mathrm}

\begin{document}

\title{Measuring dark matter-neutrino relative velocity on cosmological scales}
%\title{Measuring neutrino masses from anisotropic dark matter-neutrino correlation}
%\title{Measuring Neutrino Mass from Relative-Velocity Reconstruction}% Force line breaks with \\
%\title{Detecting Neutrino Mass from Galaxy Clustering and Lensing Cross Correlation}

\author{Hong-Ming Zhu}
\affiliation{%
Berkeley Center for Cosmological Physics and Department of Physics,
University of California, Berkeley, California 94720, USA \\
Lawrence Berkeley National Laboratory, 1 Cyclotron Road, Berkeley, California 94720, USA
% Authors' institution and/or address\\
% This line break forced with \textbackslash\textbackslash
}%

\author{Emanuele Castorina}%
\affiliation{%
Berkeley Center for Cosmological Physics and Department of Physics,
University of California, Berkeley, California 94720, USA \\
Lawrence Berkeley National Laboratory, 1 Cyclotron Road, Berkeley, California 94720, USA
% Authors' institution and/or address\\
% This line break forced with \textbackslash\textbackslash
}%

\date{\today}

\begin{abstract}
We present a new method to measure neutrino masses using the dark matter-neutrino relative velocity. 
The relative motion between dark matter and neutrinos results in a parity-odd bispectrum which can be measured from cross-correlation of different cosmic fields.
This new method is not affected by most systematics which are parity even and not limited by the knowledge of optical depth to the cosmic microwave background.
We estimate the detectability of the relative velocity effect and find that the minimal sum of neutrino masses could be detected at high significance with upcoming surveys.
\end{abstract}

\maketitle

\section{Introduction}
Neutrino mass is one of the unsolved problems of fundamental physics.
While the mass square splittings have been measured by solar neutrino and atmospheric neutrino experiments, the absolute neutrino masses are not measured, rendering neutrinos the only particles with unknown masses in the Standard Model of particle physics \cite{2018PDG}.
Many neutrino properties including mass hierarchy and chirality, i.e., whether neutrinos are Majorana or Dirac particles, 
also remain unknown.
Precision measurement of these properties can provide valuable insights
into particle physics and neutrino cosmology, as well as test possible scenarios of the early Universe \cite{2018PhysRevLett.121.251301}.

Cosmological observations can probe the sum of neutrino masses, complementary to terrestrial experiments. 
Massive neutrinos with large thermal velocity dispersion suppress the amplitude and growth rate of cosmological structures below the neutrino free-streaming scale, leading to a distinct suppression of the total matter power spectrum on small scales \cite{1980PhysRevLett.45.1980,1998PhRvL..80.5255H,2015APh....63...66A}.
Combining the primary cosmic microwave background (CMB) anisotropies and CMB lensing observations, the latest Planck Collaboration result yields a constraint on the sum of neutrino masses $\sum m_\nu< 0.12\ \mathrm{eV}$ \cite{2018arXiv180706209P}.
The next-generation surveys, e.g., DESI \cite{2016arXiv161100036D}, LSST \cite{2009arXiv0912.0201L}, CMB-S4 \cite{2016arXiv161002743A}, aim to improve the constraints to resolve the minimal mass of $\sim0.06\ \mathrm{eV}$ by mapping the distribution of mass and galaxies in the late Universe.
However, the measurement of neutrino masses from the suppression of small-scale power relative to the primary CMB is limited by the precision of optical depth to the CMB if the current measurement from the Planck satellite is not improved \cite{2016arXiv161002743A,2018PhRvD..97l3544M,2018JCAP...03..035B}.

Moreover, the forecasts of neutrino mass constraints rely on the assumption of perfect knowledge of cosmological dynamic fields, e.g., galaxy clustering and  matter distribution in the Universe. 
However, the theoretical description is not perfect and a number of theoretical errors always exist \cite{2016arXiv160200674B}.
The nonlinear dynamics between collisionless cold dark matter (CDM) and neutrinos can in principle be simulated with high precision, but the challenge is to disentangle the small neutrino effect which is a few percent level suppression of the power spectrum around nonlinear scales from the complex and poorly understood baryonic effects.
The significant degeneracy of the neutrino contribution to the power spectrum with galaxy bias parameters including higher order biases requires a much better understanding of baryonic physics to obtain a significant detection of minimal neutrino mass,
especially when the power spectrum shape information is used in the analysis \cite{2018arXiv180902120Y,2018arXiv181107636B}.
The cross-correlation of galaxy clustering and lensing has the potential to be less sensitive to systematics but it is found that the constraints only benefit mildly from the cross power spectrum \cite{2018PhRvD..97l3544M,2018arXiv181107636B,2018PhRvD..97l3540S}.
Other possible unknown systematics beyond the standard $\Lambda$CDM cosmology including nonzero curvature, dynamical dark energy, modified Einstein gravity, interactions in the dark matter sector, etc., will also affect neutrino mass constraints, where the observation of combined effects can only be ascribed to the effect of massive neutrinos. 
However, even assuming a perfect theoretical model without systematic errors, the combination of LSST and CMB-S4 can only make a $3-4\sigma$ detection of the minimal neutrino mass.
Therefore, there is clear and strong motivation to explore new effects and novel methods which can provide additional information to further improve the total signal to noise.

The principal effect of cosmic neutrinos on the large-scale structure considered by most neutrino measurement methods is caused by the relative clustering of neutrinos compared to CDM and baryons below the neutrino free-streaming scale.
Recently the relative velocity between CDM and neutrinos has been proposed as a new probe of neutrino masses \cite{2014PhRvL.113m1301Z}.
The relative advection between the initial coherent CDM and neutrino density causes a dipole contribution to the local correlation of CDM and neutrinos \cite{2014PhRvL.113m1301Z}.
On nonlinear scales, neutrinos become gravitationally focused into wakes as they flow over dark matter halos, further increasing the strength of the dipole clustering of neutrinos around high CDM density regions \cite{2016PhRvL.116n1301Z,2017MNRAS.468.2164O}.
However, the computation of dipole correlation only uses the CDM-neutrino relative velocity direction and is affected by the halo-CDM or baryon-CDM relative flows \cite{2017PhysRevD.95.083518}.
In this paper, we propose a new method to measure individual neutrino masses from the effect of CDM-neutrino relative velocity and estimate the detectability of the neutrino signal with upcoming surveys.

\section{Relative velocity}
Neutrinos and CDM are expected to have a bulk relative velocity, $\Vec{v}_{\nu c}=\Vec{v}_\nu -\Vec{v}_c $ and $\Vec{v}_{\nu c}=-\Vec{v}_{c\nu}$, due to the free streaming of neutrinos over large scales reducing the bulk motion of neutrinos. 
The relative displacement between neutrinos and CDM is thus 
\begin{equation}
    \vec{\psi}_{\nu c}(\tau)=\int_0^\tau d\tau'\vec{v}_{\nu c}(\tau'),
\end{equation}
which is an integral of velocity over the conformal time.

We can define the CDM coordinates $\vec{x}_c$ and neutrino coordinates $\vec{x}_\nu$, related by the displacement $\vec{\psi}_{c\nu}$, 
\begin{equation}
\label{eq:mapping}
    \vec{x}_c=\vec{x}_\nu + \vec{\psi}_{c\nu},
\end{equation}
in analogy to the Lagrangian space $\vec{q}$ and Eulerian space $\vec{x}$ in Lagrangian perturbation theory. 
Note that the displacement $\vec{\psi}_{c\nu}$ is the effective displacement of the CDM and neutrino fluid elements instead of the real displacement of CDM and neutrino particles.

The neutrino-CDM relative displacement leads to the advection of initially concurrent CDM and neutrino densities, reducing the correlation between these two fields, which is the same principle as the decorrelation of the Eulerian nonlinear displaced field with the Lagrangian linear initial conditions due to large infrared displacements (see Ref. \cite{2019PhRvD.100d3514S} for a recent discussion).
Therefore, the CDM and neutrino densities are not completely in phase below the free-streaming scale.

\begin{figure}[tbp]
\begin{center}
\includegraphics[width=0.48\textwidth]{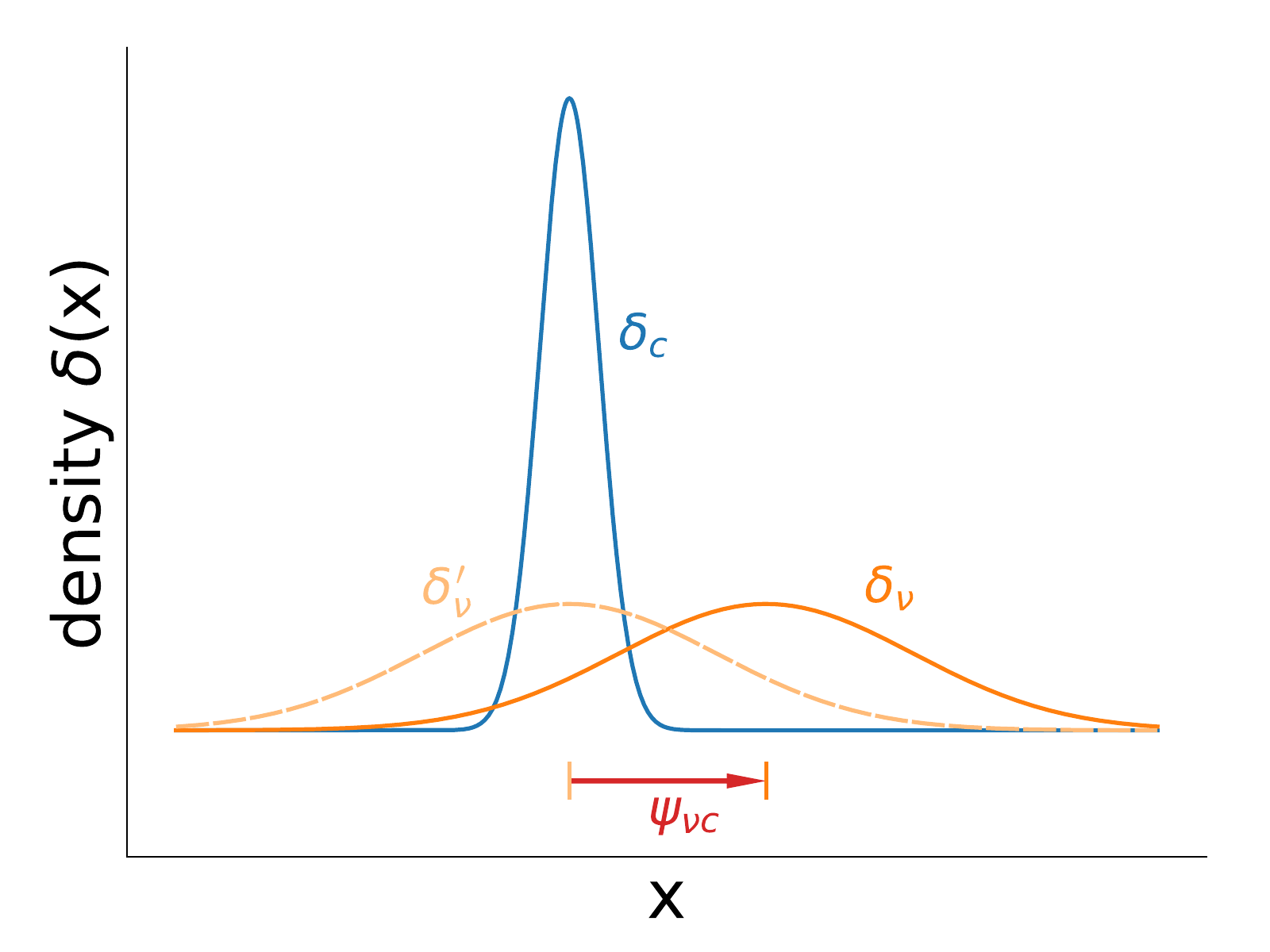}
%\hspace{-0.5cm}
%\includegraphics[width=0.24\textwidth]{./fb.pdf}
\end{center}
\vspace{-0.7cm}
\caption{A schematic picture for the CDM-neutrino relative flow. 
%Convolution of the CDM density $\delta_c$ with a Gaussian window gives the density $\delta_\nu'$, and a similar relation applies to the density $\delta_c'$ and the neutrino density $\delta_\nu$.
The observable CDM density $\delta_c$ and neutrino density $\delta_\nu$ are displaced by $\vec{\psi}_{c\nu}$ due to the relative motion of CDM and neutrino fluids.
}
\label{fig:sy}
\end{figure}

%Figure \ref{fig:sy} shows a intuitive picture of the CDM-neutrino relative flow.
%The observable CDM density $\delta_c$ and neutrino density $\delta_\nu$ are related by $\vec{\psi}_{c\nu}$.
%Smoothing the real CDM density $\delta_c$ with a Gaussian window gives the density $\delta_\nu'$, which is still fully correlated with $\delta_c$ and has a smaller power spectrum amplitude, while deconvolving the Gaussian window from $\delta_\nu$ gives the density $\delta_c'$.

An intuitive picture of the relative flow effect is that when neutrinos flow over large-scale structures, they are gravitationally focused into wakes which induces downstream overdensities relative to the CDM distribution.
Figure \ref{fig:sy} shows a schematic picture of this effect.
Such effect can be measured using cosmological probes that are sensitive to the CDM and neutrino distribution differently.

We wish to relate the CDM density field $\rho_c(\vec{x}_c)$ to its density $\rho_c'(\vec{x}_\nu)$ in the neutrino coordinates or the neutrino rest frame. From the mass conservation, we have
\begin{equation}
\label{eq:mass}
    \rho_c(\vec{x}_c)d^3\vec{x}_c=\rho_c'(\vec{x}_\nu)d^3\vec{x}_\nu.
\end{equation}
Combining the above equations and keeping only the leading order terms, we find that 
\begin{equation}
    \delta_c(\vec{x}_c)=\delta_c'(\vec{x}_c)+\vec{\psi}_{\nu c}\cdot\nabla\delta_c'(\vec{x}_c)+\nabla\cdot\vec{\psi}_{\nu c},
\end{equation}
where the $\vec{\psi}_{\nu c}\cdot\nabla\delta_c'$ term comes from the mapping of the CDM overdensity from $\vec{x}_\nu$ to $\vec{x}_c$ and the $\nabla\cdot\vec{\psi}_{\nu c}$ term comes from the Jacobian $|d^3 \vec{x}_\nu/d^3 \vec{x}_c|$. 
The CDM density field  written in neutrino coordinates $\delta_c'(\vec{x}_\nu)$ is completely in phase with the neutrino density $\delta_\nu(\vec{x}_\nu)$ with unity correlation coefficient on all scales.
Equivalently, we have
\begin{equation}
\label{eq:delta_nu}
    \delta_\nu(\vec{x}_\nu)=\delta_\nu'(\vec{x}_\nu)+\vec{\psi}_{c\nu}\cdot\nabla\delta_\nu'(\vec{x}_\nu)+\nabla\cdot\vec{\psi_{c\nu}},
\end{equation}
where the neutrino density field in the CDM rest frame $\delta_\nu'$ is completely correlated with the CDM density $\delta_c$.
We denote by prime the corresponding density fields in the other coordinates.
We can use either the CDM or the neutrino rest frame to compute the observable quantities as long as we express the fields in the same coordinate frame.
As we see, the relative bulk flow of the neutrino and CDM fluids leads to asymmetric matter distribution in the total matter density field along the direction of the CDM-neutrino relative velocity \cite{2014PhRvL.113m1301Z,2016PhRvL.116n1301Z,2017PhysRevD.95.083518,2017MNRAS.468.2164O}.

\begin{figure}[tbp]
\begin{center}
\includegraphics[width=0.48\textwidth]{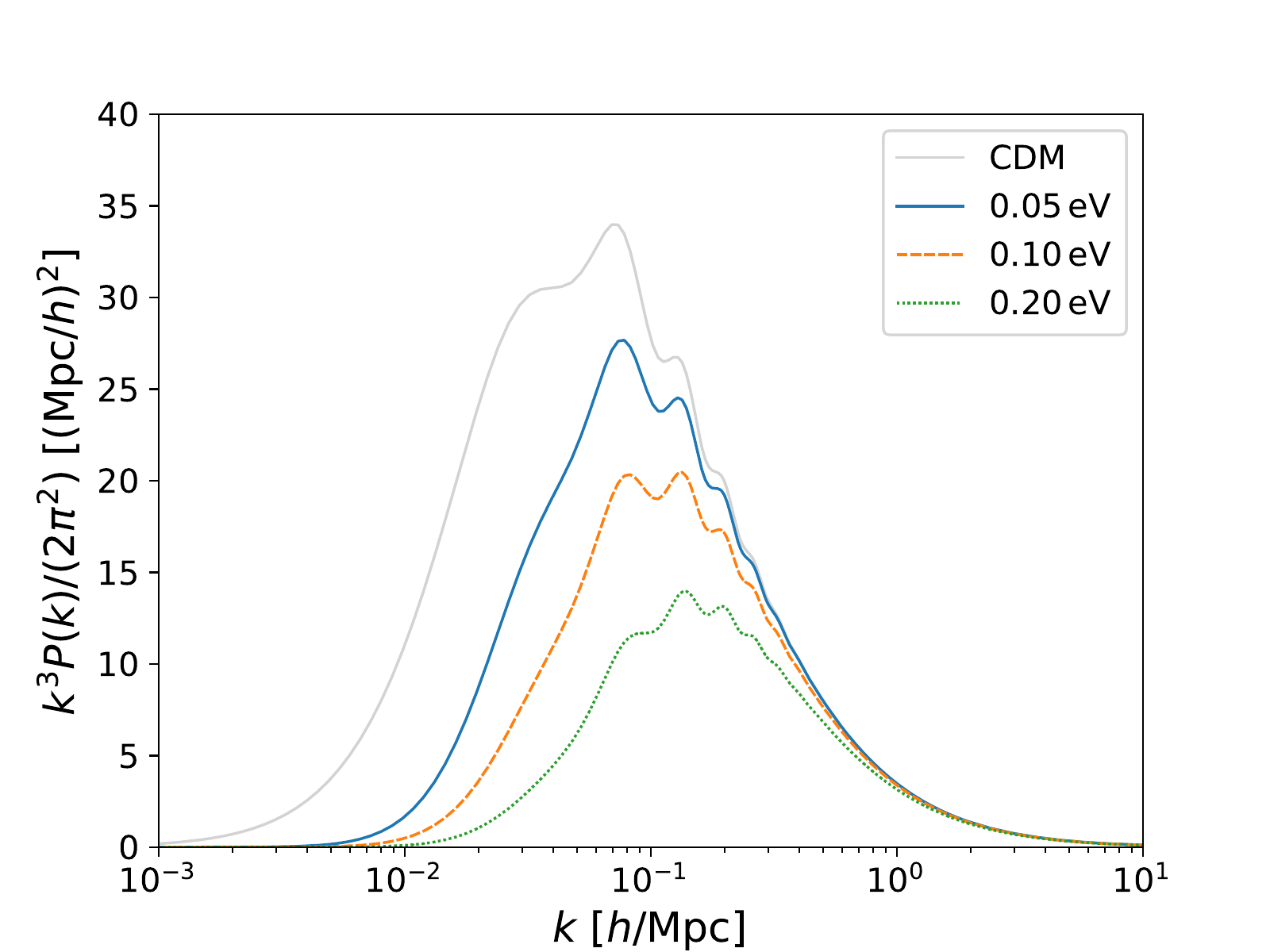}
\end{center}
\vspace{-0.7cm}
\caption{The power spectra of relative displacement fields for different neutrino masses. Neutrinos with larger masses have less power on larger scales. The light gray curve shows the power spectrum for the full CDM displacement.}
\label{fig:pk}
\end{figure}

The displacement field is dominated by the linear contribution. Then the relative displacement is given by
\begin{equation}
    \vec{\psi}_{c\nu}=\nabla\phi_{c\nu},
\end{equation}
where $\phi_{c\nu}$ is the scalar potential. 
Here the CDM-neutrino relative displacement field is basically a high-pass filtered linear CDM displacement field since neutrinos move with CDM on large scales \cite{2015PhRvD..92b3502I}.
We plot the power spectrum of $\vec{\psi}_{c\nu}$ for different neutrino masses in Fig. \ref{fig:pk}.
As $\vec{\psi}_{c\nu}$ should be much smaller than the full CDM displacement, we can treat $\vec{\psi}_{c\nu}$ perturbatively in the computation.

From Eq. (\ref{eq:delta_nu}), we see that in the CDM frame %, we see that
the relative flow introduces additional shift nonlinearities in the neutrino density field $\delta_\nu'$, which is completely correlated with the CDM density $\delta_c$.
This extra shift term $\vec{\psi}_{c\nu}\cdot\nabla\delta_\nu'$ is also orthogonal to the original neutrino density $\delta_\nu'$ and hence the CDM density $\delta_c$, i.e., the cross power spectrum
\begin{equation}
    \langle\delta_\nu'(\vec{k})(\vec{\psi}_{c\nu}\cdot\nabla\delta_\nu')^*(\vec{k})\rangle=0.
\end{equation}
This is because the local anisotropic matter distribution vanishes by parity in the averaged cross power spectrum.
%This is because the nonzero contribution to cross power comes from the connected part of $\langle\vec{\psi}\delta\delta'\rangle$ that is very small since the nonlinear contribution to displacement is negligible although here densities are both striking nonlinear fields.
%Even the connected part vanishes by parity in the averaged cross power spectrum.
The relative velocity contributes to the auto power spectrum as $\langle(\vec{\psi}\delta)^2\rangle$, which is also a second order effect. 
Therefore, we find that at linear order the effect of relative velocity vanishes at the power spectrum level.

We now consider the off-diagonal part of the covariance of $\delta_c(\vec{k})$ and $\delta_\nu(\vec{k}')$, where the diagonal part is just the cross power spectrum.
The Fourier transform of the CDM density field is
\begin{equation}
    \delta_c(\vec{k})
    =\int d^3\vec{x}_\nu(1+\delta_c'(\vec{x}_\nu))e^{-\vec{k}\cdot(\vec{x}_\nu +\vec{\psi}_{c\nu})},
\end{equation}
where $\vec{k}\neq0$.
Correlating with $\delta_\nu(\vec{k}')$, we obtain
\begin{equation}
    \langle\delta_c(\vec{k})\delta_\nu(\vec{k}')\rangle=\vec{k}\cdot\vec{K}P_{\delta_c\delta_\nu'}(k)\phi_{c\nu}(\vec{K}),
\end{equation}
where $\vec{k}+\vec{k}'=\vec{K}$ and $\vec{K}\neq0$.
In Fourier space, the effect of relative bulk flow manifests itself as correlations between modes with $\vec{k}\neq\vec{k}'$,
where the CDM-neutrino cross correlation couples to a longitudinal vector field $\nabla\phi_{c\nu}$.
A key point is that the coupling coefficient is antisymmetric in the exchange of $\delta_c$ and $\delta_\nu$; i.e.,
if we instead use the CDM coordinates, we obtain
\begin{equation}
    \langle\delta_\nu(\vec{k})\delta_c(\vec{k}')\rangle=-\vec{k}\cdot\vec{K}P_{\delta_c'\delta_\nu}(k)\phi_{c\nu}(\vec{K}).
\end{equation}
Thus the off-diagonal part of the covariance is antisymmetric, i.e., $\langle\delta_c(\vec{k}_i)\delta_\nu(\vec{k}_j)\rangle=-\langle\delta_c(\vec{k}_j)\delta_\nu(\vec{k}_i)\rangle$, because the effect of the relative velocity is parity odd under reflection in the relative velocity direction. 
Note that the Jacobian term $\nabla\cdot\vec{\psi}_{c\nu}$ only affects the diagonal part 
%of $\langle\delta_c(\vec{k}_i)\delta_\nu(\vec{k}_j)\rangle$ 
and does  not contribute to the off-diagonal signal.

We find that the relative flow between CDM and neutrinos causes an extra shift term in the density fields which otherwise will have the same phases on all scales.
In Fourier space, simply scaling the CDM density $\delta_c(\vec{k})$ by the linear transfer function $T_\nu(k)/T_c(k)$ gives the neutrino density $\delta_\nu'(\vec{k})=\delta_c(\vec{k})T_\nu(k)/T_c(k)$, which still differs from the real neutrino density field $\delta_\nu(\vec{k})$ by a phase factor depending on the local values of the CDM-neutrino relative displacement.
This is a new effect of neutrinos on the large-scale structure of the Universe other than the suppression in the amplitude and the growth rate of matter perturbations, while most current studies focus on the power spectrum amplitude and usually assume that neutrinos and CDM are completely in phase \cite{2018JCAP...03..049L,2018arXiv181011784Y,2019PhRvL.122d1302C,2019JCAP...01..059B}.

Because the relative displacement between the baryon and CDM fluids is very small, baryons and CDM should be fully correlated to much smaller scales where the baryonic processes become important.
We thus expect this relative velocity effect to be much less sensitive to the effects of baryons than the
small-scale power spectrum.
The orthogonality of the new shift term to the original CDM and neutrino density fields and the antisymmetrical property make it possible to separate the neutrino effect from other non-Gaussian signals \cite{2018JCAP...07..046F,2018PhRvD..97l3539S}.

\section{Cross correlation}
To measure the effect of relative displacement between neutrinos and dark matter, we need tracers to be sensitive to neutrinos and dark matter differently. 
Galaxies are known to be biased to the dark matter overdensity,
\begin{equation}
    \delta_g=b\delta_c+\epsilon_g,
\end{equation}
where $b$ is the galaxy bias and $\epsilon_g$ is the stochastic noise term \cite{2015JCAP...07..043C,2014JCAP...02..049C}.
This is an approximation for the galaxy response with massive neutrinos, but accurate enough for the scales considered here \cite{2018PhRvD..98d3503M,2019PhRvL.122d1302C}.
Gravitational lensing and peculiar velocities are sensitive to the total matter field
\begin{equation}
\delta_m = f_c\delta_c + f_\nu\delta_\nu'+f_\nu\vec{\psi}_{c\nu}\cdot\nabla\delta_\nu'+\epsilon_m,
\end{equation}
where $f_c=\Omega_c/(\Omega_c+\Omega_\nu)$, $f_\nu=\Omega_\nu/(\Omega_c+\Omega_\nu)$, and $\epsilon_m$ is the noise term for the corresponding cosmological probe of the total matter field and we rewrite the neutrino density using the CDM coordinates.
Note that we have assumed that baryons are part of the CDM density.

The relative velocity induces a bispectrum of the form $\langle\vec{\psi}(\vec{K})\delta(\vec{k}')\delta(\vec{k})\rangle$. Thus we can build a quadratic estimator for the relative displacement field $\vec{\psi}_{c\nu}(\vec{K})$ by summing over pairs of $\langle\delta_g(\vec{k})\delta_m(\vec{k}')\rangle$ in analogy to CMB lensing.
Here the density modes are strongly nonlinear and have a large contribution $\langle\delta(\vec{K})\delta(\vec{k})\delta(\vec{k}')\rangle$ to the bispectrum, which is symmetric in $\vec{k}$, $\vec{k}'$ and parity even in $\vec{K}$.
Since the relative displacement bispectrum is antisymmetric in $\vec{k}$, $\vec{k}'$ and parity odd in $\vec{K}$, it is orthogonal to bispectra produced by other nonlinearities. 
The parity-even signals can be removed using the antisymmetric combination of schematic form $\langle\delta_g(\vec{k})\delta_m(\vec{k}')-\delta_m(\vec{k})\delta_g(\vec{k}')\rangle$.
%Because the two fields have very correlated signals $b\delta_c$ and $f_c\delta_c+f_\nu\delta_\nu'$,
%we can also orthogonalize $\delta_m$ with respect to $\delta_g$ and obtain
%\begin{equation}
%    \delta_m^\perp=f_\nu\vec{\psi}_{c\nu}\cdot\nabla\delta_\nu'+\epsilon_m^\perp,
%\end{equation}
%where the stochastic term is about $\epsilon_m^\perp=\epsilon_m-\epsilon_g/b$, such that the parity-even nonlinearities are removed \cite{2018arXiv181110640S}. 

We can also infer the neutrino effect from galaxies of different types. 
While the relation between halos and CDM depends on the halo mass, the neutrino effect from gravitational interaction is insensitive to halo mass due to the equivalence principle. 
Neutrinos cluster on large scales and enhance the clustering of total matter.
The part of neutrino density which is out of phase with CDM density leads to a shift term in the halo density, which is about $f_\nu\vec{\psi}_{c\nu}\cdot\nabla\delta_\nu'$ up to a coefficient of order 1 \cite{2017MNRAS.468.2164O}.
If the fractional size of the neutrino effect and CDM overdensity is different between two halo fields, we can combine them in a similar way as $\delta_g$ and $\delta_m$ such that parity-even systematics cancel out.
%the neutrino effect and CDM density are orthogonal to each other \cite{2018arXiv181110640S}.
%For simplicity, we refer $\delta_g$ as tracers of the CDM overdensity $\delta_c$ and denote by $\delta_m^\perp$ the neutrino shift field after orthogonalization. 
For simplicity, we take $\delta_g$ and $\delta_m$ as an example, while the results can be directly translated to the correlation of two types of galaxies.

The quadratic estimator can be written in the  form
\begin{equation}
    \hat{\psi}_{c\nu}(\vec{K})=\frac{N_{\psi_{c\nu}}(K)}{K}\int\frac{d^3\vec{k}}{(2\pi)^3}W(\vec{k},\vec{k}')\delta_g(\vec{k})\delta_m(\vec{k}'),
\end{equation}
where $\vec{k}'=\vec{K}-\vec{k}$ \cite{2002ApJ...574..566H}. 
The weights $W(\vec{k},\vec{k}')$ can be solved by minimizing the variance of the estimator subject to the constraint $\langle\hat{\psi}_{c\nu}(\vec{K})\rangle=f_\nu K\phi_{c\nu}(\vec{K})$.
Note that the weights are parity odd such that the neutrino signals add coherently and parity-even systematics cancel out.
The noise power spectrum of the quadratic estimator is given by
%\begin{equation}
%      N_{\psi_{c\nu}}(K)=K^2\bigg[\int\frac{d^3\vec{k}}{(2\pi)^3}\frac{(b\vec{k}\cdot\vec{K}P_{\delta_c\delta_\nu'}(k))^2}{P_{\delta_g}(k)P_{\delta_m^\perp}(k')}\bigg]^{-1}.
%\end{equation}
\begin{equation}
      N_{\psi_{c\nu}}(K)=K^2\bigg[\int\frac{d^3\vec{k}}{(2\pi)^3}W(\vec{k},\vec{k}')\vec{k}\cdot\vec{K}P_{\delta_c\delta_\nu'}(k)\bigg]^{-1}.
\end{equation}

%\begin{figure}[tbp]
%\begin{center}
%\includegraphics[width=0.48\textwidth]{./f1.pdf}
%\end{center}
%\vspace{-0.7cm}
%\caption{The power spectra of relative displacement fields for different neutrino masses. Neutrinos with larger masses have less power on larger scales. The light gray curve shows the full CDM displacement.
%\EC{Linear scales?}}
%\label{fig:pk}
%\end{figure}

In Fig. \ref{fig:pk}, we show the power spectra of relative displacement fields for different neutrino masses computed using the CLASS code \cite{2011JCAP...07..034B}.
We see that the power spectrum shape depends on the value of individual neutrino mass.
The lighter neutrinos have larger relative displacements and more power on large scales.
It is easier for heavier neutrinos to follow the CDM motion, leading to less power on large scales.
However, the amplitude of the measured relative displacement field $f_\nu\vec{\psi}_{c\nu}$ using a quadratic estimator depends on the neutrino mass hierarchy, which is related to the neutrino mass fraction $f_\nu$.
For example, the measured signal will be twice as large for an inverted neutrino mass hierarchy than the normal hierarchy and will enhance the possibility of detection although the normal and inverted mass hierarchies have similar individual neutrino masses.

\section{Observability}
To measure the small neutrino effect, we need to construct relative displacement templates for different neutrino masses, including both the amplitude and phase information of a field.
The CDM displacement field itself can be obtained by recently developed nonlinear reconstruction algorithms (e.g.,  \cite{2017PhRvD..96l3502Z,2017PhRvD..96b3505S,2017JCAP...12..009S}).
The neutrino displacement can be reconstructed by applying the transfer function which depends on the neutrino mass \cite{2015PhRvD..92b3502I}. 
Both displacements $\vec{\psi}_c$, $\vec{\psi}_\nu$ are defined on a uniform Lagrangian grid $\vec{q}$.
To obtain the relative displacement $\vec{\psi}_{c\nu}$ in Eq. (\ref{eq:mapping}), we need to assign $\vec{\psi}_c-\vec{\psi}_\nu$ to the neutrino coordinates $x_\nu=\vec{q}+\vec{\psi}_\nu$ instead of the initial Lagrangian coordinates $\vec{q}$.
Correlating the relative displacement field templates with the signal measured using a quadratic estimator, we are actually constructing the optimal bispectrum estimator of $\langle\vec{\psi}_{c\nu}\delta_c\delta_\nu\rangle$ \cite{2018arXiv181013423S}.

We can perform a matched-filtering search using these relative displacement templates with different fiducial neutrino masses.
The total signal-to-noise ratio reaches its maximum with the optimal relative-displacement template that gives the best-fit individual neutrino mass.
The significance of detection is quantified by the total signal-to-noise ratio,
\begin{equation}
    \mathrm{SNR}^2=V\int\frac{d^3\vec{k}}{(2\pi)^2}\frac{f^2_\nu P^2_{{\psi}_{c\nu}}(k)}{P_{{\psi}_{c\nu}}(k)N_{\psi_{c\nu}}(k)},
\end{equation}
where $V$ is the survey volume.
In any detection of this parity-odd bispectrum, the uncertainty in the neutrino mass fraction $f_\nu$, and thus the error, is proportionate to the significance of detection; i.e., for a $5\sigma$ detection, the measurement corresponds to a $20\%$ constraint on the neutrino mass fraction $f_\nu$. 
%In principle, we can measure both the individual neutrino mass and the sum of neutrino masses using the relative velocity effect given we know the fractional size of neutrino effect and CDM overdensity for a specific field.

%\begin{figure}[tbp]
%\begin{center}
%\includegraphics[width=0.48\textwidth]{./f2.pdf}
%\end{center}
%\vspace{-0.7cm}
%\caption{The forecasted $5\sigma$ detection requirement for different neutrino masses. The long-dashed lines show $\bar{n}V=10^6,3\times10^7,10^9$ from left to right, respectively.
%}
%\label{fig:nV}
%\end{figure}

\begin{table}
\centering
\caption{
The forecasted requirement of $5\sigma$ detection for different neutrino masses. 
%The forecasted error on neutrino mass with a survey of  $V_s=1.0h^{-3}\mathrm{Gpc}^3$, %$n_g=2.4\times10^{-2}h^3\mathrm{Mpc}^{-3}$ and with current survey data, modeled with SDSS and 2dF as $V_s=0.2h^{-3}\mathrm{Gpc}^3$, $n_g V_s=1\times10^{6}$. 
%Note that substantial uncertainties exist due to unknown galaxy neutrino bias, which is a nuisance parameter that we marginalize over.
}
\label{tab:nV} 
\begin{tabular}{lc}
\hline
\hline
$\sum m_\nu$ & $V/(P_{\epsilon_m}+1/(b^2\bar{n}))$ or $V/(b_1^2/\bar{n}_2+b_2^2/\bar{n}_1)\Delta b^2$  \\
\hline
$0.05\,\mr{eV}$ & $3.6\times10^8$ \\
$0.05\,\mr{eV}\times2$  & $9\times10^7$ \\
$0.10\,\mr{eV}\times3$ & $1.5\times10^6$ \\
\hline
\hline
\end{tabular}
\end{table}

Future surveys will map the observable Universe with high precision, resolving the distribution of matter and galaxies down to nonlinear scales of $k\sim1\ h\mr{Mpc}^{-1}$.
%The observed CDM density field is dominated by cosmic variance instead of shot noise.
The ability to constrain neutrino masses relies on the noise level of the neutrino density field or the neutrino shift term which is of order $\sim f_\nu\delta_\nu$. 
For the minimal neutrino mass $\sim0.06\,\mr{eV}$, the neutrino mass fraction $f_\nu$ is about $5\times10^{-3}$. 
The observed signal depends on the neutrino and CDM biases of the fields if using two different populations of galaxies.
These require more detailed studies using precision neutrino simulations.
The total signal to noise saturates beyond linear scales as most power of the displacement field comes from large scales and noise dominates over the signal on small scales.
%For sensitivity estimation, we model the noise level of neutrino field as $P_{\epsilon_m^\perp}=1/\Bar{n}$, where $\Bar{n}$ is the number density of a survey.

Table \ref{tab:nV} shows the requirement of $5\sigma$ detection for different neutrino masses.
%as a function of survey volume and noise level of the neutrino shift term field.
%For comparison, we have $\bar{n}V=10^6$, $3\times10^7$, $10^9$, roughly corresponding to SDSS, DESI experiment and Billion Object Apparatus \cite{2016arXiv160407626D}, respectively.
For the normal hierarchy, we need $V/(P_{\epsilon_m}+1/(b^2\bar{n}))\sim3.6\times10^8$, which can be achieved with a cosmic variance dominated measurement of total matter field to scale $k\sim1\ h\mr{Mpc}^{-1}$ with $V\sim(3\ \mr{Gpc}/h)^3$ and a survey of $\sim10^9$ galaxies like the Billion Object Apparatus \cite{2016arXiv160407626D}.
If we are correlating two luminous tracers with the same response to the neutrino effect but different biases relative to the CDM density field, we need $V/(b_1^2/\bar{n}_2+b_2^2/\bar{n}_1)\Delta b^2\sim3.6\times10^8$ to have a $5\sigma$ detection, where $b_1$ and $b_2$ are biases for different types of tracers, $\bar{n}_1$ and $\bar{n}_2$ are number densities for two tracers, and $\Delta b=b_2-b_1$ is the relative bias.
%Assume $b_1\sim1.8$ for a galaxy survey and $b_2\sim0.7$ for 
The two types of galaxies could be an optical survey with the Billion Object Apparatus and a Stage-II Hydrogen Intensity Mapping experiment \cite{2018arXiv181009572C}.
The signal would be twice as large for the inverted hierarchy, which only needs to reach $\sim9\times10^7$.
For the quasidegenerate case with $\sum m_\nu=0.10\,\mr{eV}\times3$, we have $\sim1.5\times10^6$, which is already possible with current surveys.
For comparison, the number densities for the SDSS, DESI experiment, and Billion Object Apparatus are about $10^6$, $3\times10^7$, and $10^9$, respectively.

%For DESI with $\bar{n}V\sim3\times10^7$, we can already obtain a constraint $\sigma(m_\nu)=35\,\mr{meV}$, which is slightly better than the optimistic DESI forecasted constraints on neutrino masses but exploits different sources of information in the observed cosmological fields.
%and is not limited by the optical depth information.
%The numbers presented here can also be interpreted as noise for total matter field.
%In this case, 
%we need a cosmic variance dominated measurement of total matter field to scale $k\sim1\ h\mr{Mpc}^{-1}$ with $V\sim(3\ \mr{Gpc}/h)^3$, 
%which can be achieved by the combination of weak lensing measurement from LSST and peculiar velocities probed by CMB-S4 \cite{2018arXiv181013423S}.

\section{Discussion}
The relative velocity is a new effect from the neutrino free streaming over large scales, providing information orthogonal to the relative clustering effect.
%The neutrino mass measurement method proposed here differs from the one based on small scale power suppression.
In the approach based on power suppression, if baryon physics, scale-dependent bias, and other previously mentioned systematics induce a variation of the small-scale power spectrum at the percent level, it can completely swamp the neutrino signal.
In the new method using relative velocity, these effects including optical depth will only change the measured bispectrum also at the percent level.
The impact on the inferred neutrino mass would only be proportionate to any such changes, unlike for power spectrum measurements where any uncertainty is amplified by 2 orders of magnitude or more.
%The parity-odd nature also makes it possible to be separated from other parity-even non-Gaussian signals \cite{2018JCAP...07..046F,2018PhRvD..97l3539S}.
%The relative motion between CDM and neutrinos induces a parity-odd bispectrum which is less affected by other nonlinear processes and therefore can be separated from other parity-even non-Gaussian signals including nonlinearities, baryonic effects, etc \cite{2018JCAP...07..046F,2018PhRvD..97l3539S}.
%Measuring the relative velocity effect is also not limited by the accuracy of optical depth which makes it a promising probe of neutrino masses with future surveys.

The baryon-CDM relative velocity and galaxy velocity bias can generate similar features in the cross-correlation of cosmological fields \cite{2010PhRvD..82h3520T,2010JCAP...11..007D,2016PhRvL.116l1303B,2016PhRvD..94f3508S,2019arXiv190300437C,2018arXiv181202148I,2018ApJ...861...58C}.
However, the velocity bias is severely constrained by the equivalence principle \cite{2018ApJ...861...58C}. 
On scales where neutrinos flow relative to the underlying matter field, the relative velocity of baryons and CDM is expected to be small.
The large-scale baryon-CDM relative velocity can be predicted using perturbation theory and marginalized over \cite{2016PhRvL.116l1303B,2016PhRvD..94f3508S,2019arXiv190300437C}.
On very small scales where baryon physics becomes important, we need to rely on hydrodynamical simulations, but the small-scale baryon-CDM relative velocity should not correlate with the neutrino-CDM velocity as it originates from nonlinear feedback processes \cite{2018arXiv181202148I}. 
However, measuring the CDM-neutrino dipole signal only uses the direction of the relative velocity; therefore the measured dipole signal also includes the halo-CDM and baryon-CDM dipole which is degenerate with the small neutrino effect \cite{2017PhysRevD.95.083518}.
Thus, the dipole method is quite sensitive to the small-scale halo velocity bias and nonlinear baryonic physics.

%The latter would however affect measurements of the CDM-neutrino dipole relying solely on the direction of the relative velocity \cite{2017PhysRevD.95.083518}, and which are therefore sensitive to baryon physics changing the halo-CDM and baryon-CDM dipole. 
%Even the small baryon-CDM relative velocity can generate a large signal as we have much more baryons in the Universe than neutrinos.
%This large-scale feature can be estimated using hydrodynamical simulations but it is still easier compared to modeling the power spectrum to subpercent precision in the nonlinear regime.

%On the other hand, the method developed here can be used to test equivalence principle on cosmological scales and probe baryonic physics on small scales using different tracers of large-scale structure.

The detectability of the relative velocity effect depends on the fractional size of the neutrino signal and CDM contribution of a cosmic field. 
The effect of neutrinos on the matter power spectrum amplitude has been studied in detail using analytical methods and simulations, while the effect of neutrinos on the phases of different fields is less explored.
To obtain robust and unbiased constraints on neutrino masses, we need to measure the gravitational effects of nonlinear neutrino clustering for different halos with high precision $N$-body simulations including both neutrino and CDM particles 
\cite{2017RAA....17...85E,2018JCAP...09..028B,2018MNRAS.481.1486B}.
The forecasted result presented here assumes that the noises are isotropic.
A more detailed investigation about how to get enough signal to noise with various probes of the total matter field like gravitational lensing, peculiar velocities \cite{2018arXiv181013423S}, and tidal fields \cite{2016PhRvD..93j3504Z,2020JCAP...01..025V} is necessary to obtain more detailed requirements for future surveys.
%The derivation presented here is basically the Lagrangian picture of structure formation applied to nonlinear neutrino clustering which is a new direction worth exploring with rich phenomenology. 
We plan to investigate these in the future.

%In conclusion, we propose a new method to measure neutrino masses with the anisotropic CDM-neutrino correlation arising from the relative flow between CDM and neutrinos.
%The new method avoids most systematics which have significant influences on measuring small scale power spectrum and is not limited by the precision of optical depth.
%Future surveys can detect the minimal sum of neutrino masses with high significance using the new method.

\begin{acknowledgments}
We would like to thank Simone Ferraro, Derek Inman, Emmanuel Schaan, Marcel Schmittfull, Uros Seljak, and Martin White for helpful discussions.
\end{acknowledgments}

\bibliographystyle{apsrev}
\bibliography{main}
%\onecolumngrid
%\newpage

%\section{Supplemental Material}

%In the Supplemental Material, we present the details for the derivation of the optimal quadratic estimator and discuss the potential bias of the neutrino mass measurement using the relative velocity between CDM and neutrinos.

%The CDM-neutrino relative motion results in a bispectrum between the CDM density $\delta_c$, neutrino density $\delta_\nu$ and relative dispalcement $\vec{\psi}_{c\nu}$,
%\begin{equation}
%    \langle\delta_c(\vec{k})\rangle
%\end{equation}

\end{document}